\input harvmac

\input amssym

\def\unit{\relax{\rm 1\kern-.26em I}}
\def\nada{\relax{\rm 0\kern-.30em l}}
\def\tilde{\widetilde}

% \draftmode

%\def\Omega{\rho,\sigma,\nu  }

\def\det{{\rm det}}

%% MACROS
\noblackbox
\def\IL{\relax{\rm I\kern-.18em L}}
\def\IH{\relax{\rm I\kern-.18em H}}
\def\IR{\relax{\rm I\kern-.18em R}}
\def\IC{\relax\hbox{$\inbar\kern-.3em{\rm C}$}}
\def\IZ{\relax\ifmmode\mathchoice
{\hbox{\cmss Z\kern-.4em Z}}{\hbox{\cmss Z\kern-.4em Z}} {\lower.9pt\hbox{\cmsss Z\kern-.4em Z}}
{\lower1.2pt\hbox{\cmsss Z\kern-.4em Z}}\else{\cmss Z\kern-.4em Z}\fi}
\def\CM {{\cal M}}

\def\CL {{\cal L}}

\def\CO {{\cal O}}

%% MORE MACROS
\def\CM {{\cal M}}

\def\CO {{\cal O}}

\def\det{{\rm det}}

\font\manual=manfnt \def\dbend{\lower3.5pt\hbox{\manual\char127}}

\def\IZ{\relax\ifmmode\mathchoice
{\hbox{\cmss Z\kern-.4em Z}}{\hbox{\cmss Z\kern-.4em Z}} {\lower.9pt\hbox{\cmsss Z\kern-.4em Z}}
{\lower1.2pt\hbox{\cmsss Z\kern-.4em Z}}\else{\cmss Z\kern-.4em Z}\fi}
\def\half {{1\over 2}}

\def\rt2{\sqrt{2}}
\def\irt2{{1\over\sqrt{2}}}

%  \slashchar puts a slash through a character to represent contraction
%  with Dirac matrices. Use \not instead for negation of relations, and use
%  \hbar for hbar.
\def\slashchar#1{\setbox0=\hbox{$#1$}           % set a box for #1
   \dimen0=\wd0                                 % and get its size
   \setbox1=\hbox{/} \dimen1=\wd1               % get size of /
   \ifdim\dimen0>\dimen1                        % #1 is bigger
      \rlap{\hbox to \dimen0{\hfil/\hfil}}      % so center / in box
      #1                                        % and print #1
   \else                                        % / is bigger
      \rlap{\hbox to \dimen1{\hfil$#1$\hfil}}   % so center #1
      /                                         % and print /
   \fi}

\def\foursqr#1#2{{\vcenter{\vbox{
    \hrule height.#2pt
    \hbox{\vrule width.#2pt height#1pt \kern#1pt
    \vrule width.#2pt}
    \hrule height.#2pt
    \hrule height.#2pt
    \hbox{\vrule width.#2pt height#1pt \kern#1pt
    \vrule width.#2pt}
    \hrule height.#2pt
        \hrule height.#2pt
    \hbox{\vrule width.#2pt height#1pt \kern#1pt
    \vrule width.#2pt}
    \hrule height.#2pt
        \hrule height.#2pt
    \hbox{\vrule width.#2pt height#1pt \kern#1pt
    \vrule width.#2pt}
    \hrule height.#2pt}}}}
\def\psqr#1#2{{\vcenter{\vbox{\hrule height.#2pt
    \hbox{\vrule width.#2pt height#1pt \kern#1pt
    \vrule width.#2pt}
    \hrule height.#2pt \hrule height.#2pt
    \hbox{\vrule width.#2pt height#1pt \kern#1pt
    \vrule width.#2pt}
    \hrule height.#2pt}}}}
\def\sqr#1#2{{\vcenter{\vbox{\hrule height.#2pt
    \hbox{\vrule width.#2pt height#1pt \kern#1pt
    \vrule width.#2pt}
    \hrule height.#2pt}}}}
\def\square{\mathchoice\sqr65\sqr65\sqr{2.1}3\sqr{1.5}3}

\def\figin{\epsfcheck\figin}\def\figins{\epsfcheck\figins}
\def\epsfcheck{\ifx\epsfbox\UnDeFiNeD
\message{(NO epsf.tex, FIGURES WILL BE IGNORED)}
\gdef\figin##1{\vskip2in}\gdef\figins##1{\hskip.5in}% blank space instead
\else\message{(FIGURES WILL BE INCLUDED)}%
\gdef\figin##1{##1}\gdef\figins##1{##1}\fi}
\def\DefWarn#1{}
\def\figinsert{\goodbreak\midinsert}
\def\ifig#1#2#3{\DefWarn#1\xdef#1{fig.~\the\figno}
\writedef{#1\leftbracket fig.\noexpand~\the\figno}%
\figinsert\figin{\centerline{#3}}\medskip\centerline{\vbox{\baselineskip12pt \advance\hsize by
-1truein\noindent\footnotefont{\bf Fig.~\the\figno:\ } \it#2}}
\bigskip\endinsert\global\advance\figno by1}

%=====================================================================

\Title{\vbox{\baselineskip12pt\hbox{}
\hbox{WIS/01/13-FEB-DPPA}}} {\vbox{
\vskip -5cm
{\centerline{ The Effective Theory of Long Strings }}}}

\vskip  -8mm
\centerline{Ofer Aharony$^{a,b}$ and Zohar Komargodski$^{a,b}$}
\bigskip
{\sl
\centerline{$^a$Department of Particle Physics and Astrophysics}
\centerline{The Weizmann Institute of Science, Rehovot 76100, Israel}
%\centerline{\tt Ofer.Aharony@weizmann.ac.il}
\medskip
\centerline{$^b$School of Natural Sciences}\centerline{ Institute for Advanced Study, Princeton, NJ 08540, USA}
\medskip
\centerline{\tt {E-mails: Ofer.Aharony@weizmann.ac.il, Zkomargo@weizmann.ac.il}}
\bigskip
\bigskip

}

% ABSTRACT

\leftskip 8mm  \rightskip 8mm
\noindent

We present the low-energy effective theory on long strings in quantum field theory, including a streamlined review of previous literature on the subject.  Such long strings can appear in the form of solitonic strings, as in the 4d Abelian Higgs model, or in the form of confining strings, as in Yang-Mills theories. The bottom line is that upon expanding in powers of $1/L$ the energy levels of long (closed) strings (where $L$ is the length of the string),  all the terms up to (and including) order $1/L^5$ are universal. We argue that for excited strings in $D>3$ space-time dimensions there is a universal deviation at order $1/L^5$ from the naive formula that is usually used to fit lattice results. For $D=3$ this naive formula is valid even at order $1/L^5$. At order $1/L^7$ non-universal terms generically appear in all cases.
We explain the physical origin of these results, and illuminate them in three different formulations of the effective action of long strings (the relationships among which we partly clarify). In addition, we corroborate these results by an explicit computation of the effective action on long strings in confining theories which have a gravitational dual.
These predictions can be tested by precise simulations of 4d Yang-Mills theory on the lattice.
\bigskip

\leftskip 0mm  \rightskip 0mm
\Date{\hskip 8mm February 2013}

%\draftmode

\lref\LW{M.~Luscher and P.~Weisz,
  ``String excitation energies in SU(N) gauge theories beyond the  free-string
  approximation,''
  JHEP {\bf 0407}, 014 (2004)
  [arXiv:hep-th/0406205].
  %%CITATION = JHEPA,0407,014;%%
}

\lref\AK{O.~Aharony and E.~Karzbrun,
  ``On the effective action of confining strings,''
  JHEP {\bf 0906}, 012 (2009)
  [arXiv:0903.1927 [hep-th]].
  %%CITATION = JHEPA,0906,012;%%
}

%\WittenZW
\lref\WittenZW{
  E.~Witten,
  ``Anti-de Sitter space, thermal phase transition, and confinement in gauge theories,''
Adv.\ Theor.\ Math.\ Phys.\  {\bf 2}, 505 (1998).
[hep-th/9803131].
%%CITATION = hep-th/9803131%%
}

%\KlebanovHB
\lref\KlebanovHB{
  I.~R.~Klebanov and M.~J.~Strassler,
  ``Supergravity and a confining gauge theory: Duality cascades and chi SB resolution of naked singularities,''
JHEP {\bf 0008}, 052 (2000).
[hep-th/0007191].
%%CITATION = IASSNS-HEP-00-56%%
}

%\MaldacenaYY
\lref\MaldacenaYY{
  J.~M.~Maldacena and C.~Nunez,
  ``Towards the large N limit of pure N=1 superYang-Mills,''
Phys.\ Rev.\ Lett.\  {\bf 86}, 588 (2001).
[hep-th/0008001].
%%CITATION = hep-th/0008001%%
}

\lref\PS{J.~Polchinski and A.~Strominger,
  ``Effective string theory,''
  Phys.\ Rev.\ Lett.\  {\bf 67}, 1681 (1991).
  %%CITATION = PRLTA,67,1681;%%
}

%\PolyakovRD
\lref\PolyakovRD{
  A.~M.~Polyakov,
  ``Quantum geometry of bosonic strings,''
  Phys.\ Lett.\  B {\bf 103}, 207 (1981).
  %%CITATION = PHLTA,B103,207;%%
}

%\JaimungalHK
\lref\JaimungalHK{
  S.~Jaimungal, G.~W.~Semenoff and K.~Zarembo,
  ``Universality in effective strings,''
  JETP Lett.\  {\bf 69}, 509 (1999)
  [arXiv:hep-ph/9811238].
  %%CITATION = JTPLA,69,509;%%
}
%\DrummondYP
\lref\DrummondYP{
  J.~M.~Drummond,
  ``Universal subleading spectrum of effective string theory,''
  arXiv:hep-th/0411017.
  %%CITATION = HEP-TH/0411017;%%
}

%\GliozziCX
\lref\GliozziCX{
  F.~Gliozzi and M.~Meineri,
  ``Lorentz completion of effective string (and p-brane) action,''
[arXiv:1207.2912 [hep-th]].
%%CITATION = arXiv:1207.2912%%
}

%\MeineriEW
\lref\MeineriEW{
  M.~Meineri,
  ``Lorentz completion of effective string action,''
[arXiv:1301.3437 [hep-th]].
%%CITATION = arXiv:1301.3437%%
}

%\AharonyGB
\lref\AD{
  O.~Aharony and M.~Dodelson,
  ``Effective String Theory and Nonlinear Lorentz Invariance,''
JHEP {\bf 1202}, 008 (2012).
[arXiv:1111.5758 [hep-th]].
%%CITATION = arXiv:1111.5758%%
}

%\DubovskySH
\lref\NYU{
  S.~Dubovsky, R.~Flauger and V.~Gorbenko,
  ``Effective String Theory Revisited,''
[arXiv:1203.1054 [hep-th]].
%%CITATION = arXiv:1203.1054%%
}

%\BigazziZE
\lref\BigazziZE{
  F.~Bigazzi, A.~L.~Cotrone, L.~Martucci and L.~A.~Pando Zayas,
  ``Wilson loop, Regge trajectory and hadron masses in a Yang-Mills theory from semiclassical strings,''
Phys.\ Rev.\ D {\bf 71}, 066002 (2005).
[hep-th/0409205].
%%CITATION = hep-th/0409205%%
}

%\DubovskyWK
\lref\DubovskyWK{
  S.~Dubovsky, R.~Flauger and V.~Gorbenko,
  ``Solving the Simplest Theory of Quantum Gravity,''
[arXiv:1205.6805 [hep-th]].
%%CITATION = arXiv:1205.6805%%
}

\lref\Talks{O.~Aharony, Z.~Komargodski, and A.~Schwimmer, presented by O.~Aharony at the Strings 2009 conference, June 2009, and at the
ECT* workshop on ``Confining flux tubes and strings'', July 2010 (talks available online).}

%\NatsuumeKY
\lref\Natsuume{
  M.~Natsuume,
  ``Nonlinear sigma model for string solitons,''
Phys.\ Rev.\ D {\bf 48}, 835 (1993).
[hep-th/9206062].
%%CITATION = hep-th/9206062%%
}

%\AharonyGA
\lref\AFK{
  O.~Aharony, M.~Field and N.~Klinghoffer,
  ``The effective string spectrum in the orthogonal gauge,''
JHEP {\bf 1204}, 048 (2012).
[arXiv:1111.5757 [hep-th]].
%%CITATION = arXiv:1111.5757%%
}

%\JaimungalHK
\lref\JaimungalHK{
  S.~Jaimungal, G.~W.~Semenoff and K.~Zarembo,
  ``Universality in effective strings,''
JETP Lett.\  {\bf 69}, 509 (1999).
[hep-ph/9811238].
%%CITATION = hep-ph/9811238%%
}

%\PolyakovCS
\lref\PolyakovCS{
  A.~M.~Polyakov,
  ``Fine Structure of Strings,''
Nucl.\ Phys.\ B {\bf 268}, 406 (1986).
}

%\KavalovDI
\lref\KavalovDI{
  A.~R.~Kavalov and A.~G.~Sedrakian,
  ``Quantum Geometry Of Covariant Superstring With N=1 Global Supersymmetry,''
Phys.\ Lett.\ B {\bf 182}, 33 (1986).
%%CITATION = EFI-815-42-85-YEREVAN%%
}

%\AharonyGB
\lref\AD{
  O.~Aharony and M.~Dodelson,
  ``Effective String Theory and Nonlinear Lorentz Invariance,''
JHEP {\bf 1202}, 008 (2012).
[arXiv:1111.5758 [hep-th]].
%%CITATION = arXiv:1111.5758%%
}

%\MeyerQX
\lref\MeyerQX{
  H.~B.~Meyer,
  ``Poincare invariance in effective string theories,''
JHEP {\bf 0605}, 066 (2006).
[hep-th/0602281].
%%CITATION = hep-th/0602281%%
}

%\AharonyCX
\lref\AharonyCX{
  O.~Aharony and M.~Field,
  ``On the effective theory of long open strings,''
JHEP {\bf 1101}, 065 (2011).
[arXiv:1008.2636 [hep-th]].
%%CITATION = arXiv:1008.2636%%
}

%\AharonyDB
\lref\AharonyDB{
  O.~Aharony and N.~Klinghoffer,
  ``Corrections to Nambu-Goto energy levels from the effective string action,''
JHEP {\bf 1012}, 058 (2010).
[arXiv:1008.2648 [hep-th]].
%%CITATION = arXiv:1008.2648%%
}

%\TeperUF
\lref\TeperUF{
  M.~Teper,
  ``Large N and confining flux tubes as strings - a view from the lattice,''
Acta Phys.\ Polon.\ B {\bf 40}, 3249 (2009).
[arXiv:0912.3339 [hep-lat]].
%%CITATION = arXiv:0912.3339%%
}

%\AthenodorouCS
\lref\AthenodorouCS{
  A.~Athenodorou, B.~Bringoltz and M.~Teper,
  ``Closed flux tubes and their string description in D=3+1 SU(N) gauge theories,''
JHEP {\bf 1102}, 030 (2011).
[arXiv:1007.4720 [hep-lat]].
%%CITATION = arXiv:1007.4720%%
}

%\AthenodorouRX
\lref\AthenodorouRX{
  A.~Athenodorou, B.~Bringoltz and M.~Teper,
  ``Closed flux tubes and their string description in D=2+1 SU(N) gauge theories,''
JHEP {\bf 1105}, 042 (2011).
[arXiv:1103.5854 [hep-lat]].
%%CITATION = arXiv:1103.5854%%
}

%\BrandtTC
\lref\BrandtTC{
  B.~B.~Brandt and P.~Majumdar,
  ``Spectrum of the QCD flux tube in 3d SU(2) lattice gauge theory,''
Phys.\ Lett.\ B {\bf 682}, 253 (2009).
[arXiv:0905.4195 [hep-lat]].
%%CITATION = arXiv:0905.4195%%
}

%\BrandtBW
\lref\BrandtBW{
  B.~B.~Brandt,
  ``Probing boundary-corrections to Nambu-Goto open string energy levels in 3d SU(2) gauge theory,''
JHEP {\bf 1102}, 040 (2011).
[arXiv:1010.3625 [hep-lat]].
%%CITATION = arXiv:1010.3625%%
}

%\KavalovNX
\lref\KavalovNX{
  A.~R.~Kavalov, I.~K.~Kostov and A.~G.~Sedrakian,
  ``Dirac And Weyl Fermion Dynamics On Two-dimensional Surface,''
Phys.\ Lett.\ B {\bf 175}, 331 (1986).
%%CITATION = EFI-895-46-86-YEREVAN%%
}

%\MazurNR
\lref\MazurNR{
  P.~O.~Mazur and V.~P.~Nair,
  ``Strings in QCD and theta vacua,''
Nucl.\ Phys.\ B {\bf 284}, 146 (1987).
%%CITATION = NSF-ITP-86-91%%
}

%\BalachandranEP
\lref\BalachandranEP{
  A.~P.~Balachandran, F.~Lizzi and G.~Sparano,
  ``Theta Vacua, Fermions From Bosons, Solitons And Wess-zumino Terms In String Models,''
Nucl.\ Phys.\ B {\bf 263}, 608 (1986)..
%%CITATION = SU-4222-315%%
}

%\KolFQ
\lref\KolFQ{
  U.~Kol and J.~Sonnenschein,
  ``Can holography reproduce the QCD Wilson line?,''
JHEP {\bf 1105}, 111 (2011).
[arXiv:1012.5974 [hep-th]].
%%CITATION = arXiv:1012.5974%%
}

%\MakeenkoDQ
\lref\MakeenkoDQ{
  Y.~Makeenko,
  ``Effective String Theory and QCD Scattering Amplitudes,''
Phys.\ Rev.\ D {\bf 83}, 026007 (2011).
[arXiv:1012.0708 [hep-th]].
%%CITATION = arXiv:1012.0708%%
}

%\MakeenkoUG
\lref\MakeenkoUG{
  Y.~Makeenko,
  ``QCD String as an Effective String,''
[arXiv:1206.0922 [hep-th]].
%%CITATION = arXiv:1206.0922%%
}

%\MaldacenaRE
\lref\MaldacenaRE{
  J.~M.~Maldacena,
  ``The Large N limit of superconformal field theories and supergravity,''
Adv.\ Theor.\ Math.\ Phys.\  {\bf 2}, 231 (1998).
[hep-th/9711200].
%%CITATION = hep-th/9711200%%
}

%\GubserBC
\lref\GubserBC{
  S.~S.~Gubser, I.~R.~Klebanov and A.~M.~Polyakov,
  ``Gauge theory correlators from noncritical string theory,''
Phys.\ Lett.\ B {\bf 428}, 105 (1998).
[hep-th/9802109].
%%CITATION = hep-th/9802109%%
}

%\WittenQJ
\lref\WittenQJ{
  E.~Witten,
  ``Anti-de Sitter space and holography,''
Adv.\ Theor.\ Math.\ Phys.\  {\bf 2}, 253 (1998).
[hep-th/9802150].
%%CITATION = hep-th/9802150%%
}

%\ThiemannQU
\lref\ThiemannQU{
  T.~Thiemann,
  ``The LQG string: Loop quantum gravity quantization of string theory I: Flat target space,''
Class.\ Quant.\ Grav.\  {\bf 23}, 1923 (2006).
[hep-th/0401172].
%%CITATION = hep-th/0401172%%
}

%\MeusburgerSA
\lref\MeusburgerSA{
  C.~Meusburger and K.~-H.~Rehren,
  ``Algebraic quantization of the closed bosonic string,''
Commun.\ Math.\ Phys.\  {\bf 237}, 69 (2003).
[math-ph/0202041].
%%CITATION = math-ph/0202041%%
}

%\AharonyGA
\lref\AharonyGA{
  O.~Aharony, M.~Field and N.~Klinghoffer,
  ``The effective string spectrum in the orthogonal gauge,''
JHEP {\bf 1204}, 048 (2012).
[arXiv:1111.5757 [hep-th]].
%%CITATION = arXiv:1111.5757%%
}

%\PolyakovRD
\lref\PolyakovRD{
  A.~M.~Polyakov,
  ``Quantum Geometry of Bosonic Strings,''
Phys.\ Lett.\ B {\bf 103}, 207 (1981).
%%CITATION = Print-81-0351 (LANDAU INST)%%
}

%\ArvisFP
\lref\ArvisFP{
  J.~F.~Arvis,
  ``The Exact q Anti-q Potential In Nambu String Theory,''
Phys.\ Lett.\ B {\bf 127}, 106 (1983).
%%CITATION = LPTENS 83/6%%
}

%\ZwiebachTJ
\lref\Zwiebach{
  B.~Zwiebach,
  ``A first course in string theory,''
Cambridge, UK: Univ. Pr. (2009) 673 p.
}

%\IvanovZQ
\lref\IvanovZQ{
  E.~A.~Ivanov and V.~I.~Ogievetsky,
  ``The Inverse Higgs Phenomenon in Nonlinear Realizations,''
Teor.\ Mat.\ Fiz.\  {\bf 25}, 164 (1975)..
%%CITATION = JINR-E2-8593%%
}

%\LowBW
\lref\LowBW{
  I.~Low and A.~V.~Manohar,
  ``Spontaneously broken space-time symmetries and Goldstone's theorem,''
Phys.\ Rev.\ Lett.\  {\bf 88}, 101602 (2002).
[hep-th/0110285].
%%CITATION = hep-th/0110285%%
}

%\LohmayerAJ
\lref\LohmayerAJ{
  R.~Lohmayer and H.~Neuberger,
  ``Large-N string tension from rectangular Wilson loops,''
PoS LATTICE {\bf 2012}, 220 (2012).
[arXiv:1210.7484 [hep-lat]].
%%CITATION = arXiv:1210.7484%%
}

%\BilloFD
\lref\BilloFD{
  M.~Billo, M.~Caselle and R.~Pellegrini,
  ``New numerical results and novel effective string predictions for Wilson loops,''
JHEP {\bf 1201}, 104 (2012).
[arXiv:1107.4356 [hep-th]].
%%CITATION = arXiv:1107.4356%%
}

%\BilloDA
\lref\BilloDA{
  M.~Billo, M.~Caselle, F.~Gliozzi, M.~Meineri and R.~Pellegrini,
  ``The Lorentz-invariant boundary action of the confining string and its universal contribution to the inter-quark potential,''
JHEP {\bf 1205}, 130 (2012).
[arXiv:1202.1984 [hep-th]].
%%CITATION = arXiv:1202.1984%%
}

%\GliozziZV
\lref\GliozziZV{
  F.~Gliozzi, M.~Pepe and U.~-J.~Wiese,
  ``The Width of the Confining String in Yang-Mills Theory,''
Phys.\ Rev.\ Lett.\  {\bf 104}, 232001 (2010).
[arXiv:1002.4888 [hep-lat]].
%%CITATION = arXiv:1002.4888%%
}

%\GliozziZT
\lref\GliozziZT{
  F.~Gliozzi, M.~Pepe and U.~-J.~Wiese,
  ``The Width of the Color Flux Tube at 2-Loop Order,''
JHEP {\bf 1011}, 053 (2010).
[arXiv:1006.2252 [hep-lat]].
%%CITATION = arXiv:1006.2252%%
}

%\GliozziJH
\lref\GliozziJH{
  F.~Gliozzi, M.~Pepe and U.~-J.~Wiese,
  ``Linear Broadening of the Confining String in Yang-Mills Theory at Low Temperature,''
JHEP {\bf 1101}, 057 (2011).
[arXiv:1010.1373 [hep-lat]].
%%CITATION = arXiv:1010.1373%%
}

%\PohlmeyerSQ
\lref\PohlmeyerSQ{
  K.~Pohlmeyer,
  ``A Group Theoretical Approach To The Quantization Of The Free Relativistic Closed String,''
Phys.\ Lett.\ B {\bf 119}, 100 (1982).
%%CITATION = FREIBURG-THEP-82/3%%
}

%\BahnsWJ
\lref\BahnsWJ{
  D.~Bahns, K.~Rejzner and J.~Zahn,
  ``The effective theory of strings,''
[arXiv:1204.6263 [math-ph]].
%%CITATION = arXiv:1204.6263%%
}

%\LuscherFR
\lref\LuscherFR{
  M.~Luscher, K.~Symanzik and P.~Weisz,
  ``Anomalies of the Free Loop Wave Equation in the WKB Approximation,''
Nucl.\ Phys.\ B {\bf 173}, 365 (1980).
%%CITATION = DESY 80/31%%
}

%\LuscherAC
\lref\LuscherAC{
  M.~Luscher,
  ``Symmetry Breaking Aspects of the Roughening Transition in Gauge Theories,''
Nucl.\ Phys.\ B {\bf 180}, 317 (1981)..
%%CITATION = DESY 80/87%%
}

%\GoddardQH
\lref\GoddardQH{
  P.~Goddard, J.~Goldstone, C.~Rebbi and C.~B.~Thorn,
  ``Quantum dynamics of a massless relativistic string,''
Nucl.\ Phys.\ B {\bf 56}, 109 (1973).
}

%\GliozziHJ
\lref\GliozziHJ{
  F.~Gliozzi,
  ``Dirac-Born-Infeld action from spontaneous breakdown of Lorentz symmetry in brane-world scenarios,''
Phys.\ Rev.\ D {\bf 84}, 027702 (2011).
[arXiv:1103.5377 [hep-th]].
%%CITATION = arXiv:1103.5377%%
}

%\BotelhoZW
\lref\BotelhoZW{
  L.~C.~L.~Botelho,
  ``Covariant path integral for Nambu-Goto string theory,''
Phys.\ Rev.\ D {\bf 49}, 1975 (1994).
}

%\MerminFE
\lref\MerminFE{
  N.~D.~Mermin and H.~Wagner,
  ``Absence of ferromagnetism or antiferromagnetism in one-dimensional or two-dimensional isotropic Heisenberg models,''
Phys.\ Rev.\ Lett.\  {\bf 17}, 1133 (1966).
}

%\ColemanCI
\lref\ColemanCI{
  S.~R.~Coleman,
  ``There are no Goldstone bosons in two-dimensions,''
Commun.\ Math.\ Phys.\  {\bf 31}, 259 (1973).
}

\lref\Aharony{
  O.~Aharony, Z.~Komargodski and A.~Schwimmer,
  presented by O. Aharony at the Strings 2009 conference, June 2009,
  {\tt http://strings2009.roma2.infn.it/talks/
  Aharony\_Strings09.ppt}, and
  at the ECT* workshop on ``Confining flux tubes and strings'', July 2010,
  {\tt http://www.ect.it/Meetings/ConfsWksAndCollMeetings/
  ConfWksDocument/2010/talks/Workshop\_05\_07\_2010/Aharony.ppt}.}

%\SchwimmerZA
\lref\SchwimmerZA{
  A.~Schwimmer and S.~Theisen,
  ``Spontaneous Breaking of Conformal Invariance and Trace Anomaly Matching,''
  Nucl.\ Phys.\  B {\bf 847}, 590 (2011)
  [arXiv:1011.0696 [hep-th]].
  %%CITATION = NUPHA,B847,590;%%
}

%\KomargodskiVJ
\lref\KomargodskiVJ{
  Z.~Komargodski and A.~Schwimmer,
  ``On Renormalization Group Flows in Four Dimensions,''
JHEP {\bf 1112}, 099 (2011).
[arXiv:1107.3987 [hep-th]].
%%CITATION = arXiv:1107.3987%%
}

%\VyasWG
\lref\VyasWG{
  V.~Vyas,
  ``Intrinsic Thickness of QCD Flux-Tubes,''
[arXiv:1004.2679 [hep-th]].
%%CITATION = arXiv:1004.2679%%
}

%\VyasDG
\lref\VyasDG{
  V.~Vyas,
  ``Heavy Quark Potential from Gauge/Gravity Duality: A Large D Analysis,''
[arXiv:1209.0883 [hep-th]].
%%CITATION = arXiv:1209.0883%%
}

%\BilloIX
\lref\BilloIX{
  M.~Billo, M.~Caselle, V.~Verduci and M.~Zago,
  ``New results on the effective string corrections to the inter-quark potential,''
PoS LATTICE {\bf 2010}, 273 (2010).
[arXiv:1012.3935 [hep-lat]].
%%CITATION = arXiv:1012.3935%%
}

%\BakerXN
\lref\BakerXN{
  M.~Baker and R.~Steinke,
  ``An Effective string theory of Abrikosov-Nielsen-Olesen vortices,''
Phys.\ Lett.\ B {\bf 474}, 67 (2000).
[hep-ph/9905375].
%%CITATION = hep-ph/9905375%%
}

%\BakerCI
\lref\BakerCI{
  M.~Baker and R.~Steinke,
  ``Effective string theory of vortices and Regge trajectories,''
Phys.\ Rev.\ D {\bf 63}, 094013 (2001).
[hep-ph/0006069].
%%CITATION = hep-ph/0006069%%
}

%\GregoryQV
\lref\GregoryQV{
  R.~Gregory,
  ``Effective Action For A Cosmic String,''
Phys.\ Lett.\ B {\bf 206}, 199 (1988).
%%CITATION = DAMTP-88-1%%
}

%\OrlandQT
\lref\OrlandQT{
  P.~Orland,
  ``Extrinsic curvature dependence of Nielsen-Olesen strings,''
Nucl.\ Phys.\ B {\bf 428}, 221 (1994).
[hep-th/9404140].
%%CITATION = hep-th/9404140%%
}

%\AkhmedovMW
\lref\AkhmedovMW{
  E.~T.~Akhmedov, M.~N.~Chernodub, M.~I.~Polikarpov and M.~A.~Zubkov,
  ``Quantum theory of strings in Abelian Higgs model,''
Phys.\ Rev.\ D {\bf 53}, 2087 (1996).
[hep-th/9505070].
%%CITATION = hep-th/9505070%%
}

%\DubovskySH
\lref\NYU{
  S.~Dubovsky, R.~Flauger and V.~Gorbenko,
  ``Effective String Theory Revisited,''
[arXiv:1203.1054 [hep-th]].
%%CITATION = arXiv:1203.1054%%
}

%\MaedaPD
\lref\MaedaPD{
  K.~-i.~Maeda and N.~Turok,
  ``Finite Width Corrections to the Nambu Action for the Nielsen-Olesen String,''
Phys.\ Lett.\ B {\bf 202}, 376 (1988).
%%CITATION = FERMILAB-PUB-87-209-A%%
}

%\HananyEA
\lref\HananyEA{
  A.~Hanany and D.~Tong,
  ``Vortex strings and four-dimensional gauge dynamics,''
JHEP {\bf 0404}, 066 (2004).
[hep-th/0403158].
%%CITATION = hep-th/0403158%%
}

%\ShifmanDR
\lref\ShifmanDR{
  M.~Shifman and A.~Yung,
  ``NonAbelian string junctions as confined monopoles,''
Phys.\ Rev.\ D {\bf 70}, 045004 (2004).
[hep-th/0403149].
%%CITATION = hep-th/0403149%%
}
%\EtoPG
\lref\EtoPG{
  M.~Eto, Y.~Isozumi, M.~Nitta, K.~Ohashi and N.~Sakai,
  ``Solitons in the Higgs phase: The Moduli matrix approach,''
J.\ Phys.\ A A {\bf 39}, R315 (2006).
[hep-th/0602170].
%%CITATION = hep-th/0602170%%
}
%\TongUN
\lref\TongUN{
  D.~Tong,
  ``TASI lectures on solitons: Instantons, monopoles, vortices and kinks,''
[hep-th/0509216].
%%CITATION = hep-th/0509216%%
}
%\ShifmanCE
\lref\ShifmanCE{
  M.~Shifman and A.~Yung,
  ``Supersymmetric Solitons and How They Help Us Understand Non-Abelian Gauge Theories,''
Rev.\ Mod.\ Phys.\  {\bf 79}, 1139 (2007).
[hep-th/0703267].
%%CITATION = hep-th/0703267%%
}

%\ShifmanID
\lref\ShifmanID{
  M.~Shifman and A.~Yung,
  ``Non-Abelian Confinement in N=2 Supersymmetric QCD: Duality and Kinks on Confining Strings,''
Phys.\ Rev.\ D {\bf 81}, 085009 (2010).
[arXiv:1002.0322 [hep-th]].
%%CITATION = arXiv:1002.0322%%
}

%\HananyHP
\lref\HananyHP{
  A.~Hanany and D.~Tong,
  ``Vortices, instantons and branes,''
JHEP {\bf 0307}, 037 (2003).
[hep-th/0306150].
%%CITATION = hep-th/0306150%%
}

%\DubovskyGI
\lref\newNYU{
  S.~Dubovsky, R.~Flauger and V.~Gorbenko,
  ``Evidence for a new particle on the worldsheet of the QCD flux tube,''
[arXiv:1301.2325 [hep-th]].
%%CITATION = arXiv:1301.2325%%
}

%\EtoCV
\lref\EtoCV{
  M.~Eto, T.~Fujimori, S.~B.~Gudnason, Y.~Jiang, K.~Konishi, M.~Nitta and K.~Ohashi,
  ``Vortices and Monopoles in Mass-deformed SO and USp Gauge Theories,''
JHEP {\bf 1112}, 017 (2011).
[arXiv:1108.6124 [hep-th]].
%%CITATION = YGHP-11-45%%
}

\lref\Talks{O.~Aharony, Z.~Komargodski, and A.~Schwimmer, presented by O.~Aharony at the Strings 2009 conference, June 2009, and at the
ECT* workshop on ``Confining flux tubes and strings'', July 2010 (talks available online).}

\lref\newAK{N.~Klinghoffer, work in progress.}

%\GasserGG
\lref\GasserGG{
  J.~Gasser and H.~Leutwyler,
  ``Chiral Perturbation Theory: Expansions in the Mass of the Strange Quark,''
Nucl.\ Phys.\ B {\bf 250}, 465 (1985)..
%%CITATION = CERN-TH-3798%%
}

%\GasserYG
\lref\GasserYG{
  J.~Gasser and H.~Leutwyler,
  ``Chiral Perturbation Theory to One Loop,''
Annals Phys.\  {\bf 158}, 142 (1984)..
%%CITATION = CERN-TH-3689%%
}

%\GreenSP
\lref\GreenSP{
  M.~B.~Green, J.~H.~Schwarz and E.~Witten,
  ``Superstring Theory. Vol. 1: Introduction,''
Cambridge, Uk: Univ. Pr. ( 1987) 469 P. ( Cambridge Monographs On Mathematical Physics).
}

%\SundrumQT
\lref\SundrumQT{
  R.~Sundrum,
  ``Hadronic string from confinement,''
[hep-ph/9702306].
%%CITATION = hep-ph/9702306%%
}

\newsec{Introduction and Summary of Results}

Many quantum field theories admit long, stable, string-like objects. For example, this occurs in the 4d Abelian Higgs model (with its perturbative Abrikosov-Nielsen-Olesen strings) and also in 3d and 4d pure Yang-Mills theories (with their non-perturbative confining strings). Domain walls in 3d quantum field theories (such as the Ising model) are another example. The purpose of this work is to
summarize the universal properties
of these long strings. For simplicity, we will restrict ourselves to theories that have a mass gap in the bulk.
This is true in particular in the Abelian Higgs model and in Yang-Mills theory.\foot{Our discussion holds for any stable string; it is independent of taking any large $N$ limit in the field theory. In the large $N$ limit of non-Abelian gauge theories we expect the interactions between confining strings to become small, but this is irrelevant for the low energy effective actions we discuss. This is because there is an energy gap to creating additional strings on top of the long string we are considering.}

Even though there are no massless excitations in the bulk, in the presence of the long string there are always massless excitations confined to the string. This is easy to see, since the string breaks translational invariance and Lorentz invariance; for a straight string moving in $D$ space-time dimensions the symmetry breaking pattern\foot{This is true in classical physics. Quantum mechanically, since the string worldsheet theory is $1+1$ dimensional, there is not really a ground state of the string with broken symmetry \refs{\MerminFE,\ColemanCI}; the wavefunction of the string will always spread out in the transverse directions (unless its edges are fixed), and the actual eigenstates of the Hamiltonian have fixed transverse momentum rather than fixed transverse position. We can ignore this issue in our analysis (as in fundamental string theory), going to transverse-momentum-space only in the end when we compute the energy levels.}
is $ISO(D-1,1) \to ISO(1,1)\times SO(D-2)$.
This leads to $(D-2)$ Nambu-Goldstone bosons (NGBs). (Naively one might have expected more, but
%as is well known,
the generators of $ISO(D-1,1)$ are not all independent operators on the string worldsheet; see e.g. \refs{\IvanovZQ,\LowBW}.)

We will assume that there are no other excitations on the world-sheet besides these $(D-2)$ Nambu-Goldstone bosons.\foot{This will generically be true, unless there are additional symmetries in the problem. There are fascinating realizations of physical systems admitting more complicated theories on the worldsheet. Some of the well-known examples come from supersymmetric QCD-like theories, see~\refs{\HananyHP,\HananyEA,\ShifmanDR}, the reviews~\refs{\EtoPG,\TongUN,\ShifmanCE}, and references therein. For more recent works on the subject see, for example, \refs{\ShifmanID,\EtoCV}. It would be interesting to generalize the present analysis to such theories.}
One can then consistently describe a low-energy effective action that includes only these NGBs \SundrumQT, up to the mass of the lightest massive states on the string worldsheet or in the bulk, or of the typical energy scale of the string (which is the square root of the string tension). For example, we can prepare waves of these NGBs (i.e. wrinkles of the string) and scatter them. This is described at low enough energies using this ``effective string action.''
We can also consider strings of finite (but very large in units of the inverse mass scale of the problem) length and ask about the total energy of the ground state, and the energies of excited states. These
are well defined when the long finite string is stable; it could be, for instance, a long string wrapping a circle of circumference $L$, or an open string stretched between two end-points at distance $L$.

 In the last few years there has been dramatic progress (see e.g.~\refs{\BilloIX\BrandtTC\AthenodorouCS-\AthenodorouRX}, and the review \TeperUF\ for a more complete list of references) in measuring the energies of various states of such strings via lattice simulations. This is the main motivation for our work.
  Given a closed string of length $L$, with tension $T$, in the $n$'th excited state of the universal massless modes on the worldsheet,\foot{
  %Generally, there are also states on the world-sheet that are separated by a gap at $L\rightarrow\infty$ from the classical contribution to the energy, $TL$. For instance, they can be due to massive particles on the worldsheet. We do not discuss those states here. } 
  The energies of all these states approach the classical energy of the string, $TL$,
in the large $L$ limit. In general, there
could be also other excited states of the string, that are separated by a gap
from $TL$ as $L \to \infty$; for instance, they can be related to massive particles on the
string worldsheet. We do not discuss these states here.}
the energy of a state with zero transverse momentum assumes the general form (expanding around large $L$)
\eqn\energylevel{E_n=TL+{a^{(1)}_{n}\over L}+{a^{(2)}_{n}\over TL^3}+{a^{(3)}_{n}\over T^2L^5}\cdots~. }
The lattice techniques allow a precise determination of the coefficient $a_{0}^{(1)}$,  and in some theories (mostly in 3d) higher coefficients for the ground state and excited states were also measured. The measured leading coefficients in the large $L$ expansion seem to be model independent (they agree in all the theories simulated on the lattice so far). A simple ansatz
\eqn\energylevelsi{E_n=TL\sqrt{1+{8\pi\over TL^2}\left(n-{D-2\over 24}\right)}~}
for the energies of closed string states with no longitudinal momentum seems to fit very well all the currently known results (at least within the range of convergence of the expansion in powers of $1/L$).
The formula~\energylevelsi\ is motivated by a naive light-cone quantization~\ArvisFP\ of the simplest effective action of a long string, the Nambu-Goto action (which is just $T$ times the induced area of the worldsheet).

 We would like to understand to what extent~\energylevelsi\ can be justified on theoretical grounds, and in particular, we would like to find at what order in the expansion in $1/L$ the first deviations from this ansatz should take place, if at all.

 In some cases the effective string action can be computed analytically. This is the case for strings in weakly coupled field theories like the Abelian Higgs model
\refs{\GregoryQV\MaedaPD\OrlandQT\AkhmedovMW\BakerXN-\BakerCI}, and for strings in weakly curved holographic backgrounds \AK. But generally (in particular in
 Yang-Mills theories) it is not known how to  compute the effective string action directly, so it is useful to understand which terms in this action are universal (and, thus, can be theoretically predicted) and which terms are not (and, thus, measuring them teaches us about the underlying field theory).

This paper is based mostly on work reported in~\Aharony, but our presentation attempts to streamline and clarify it, and also to incorporate the results of various later publications such as ~\refs{\AharonyCX\AharonyDB \KolFQ\AharonyGA-\AD}, and notably~\NYU. Therefore, we hope that this work can also serve as a review of the subject.

Let us begin by stating the main results for closed confining strings\foot{Long open strings were discussed, for instance, in \refs{\LW,\AharonyCX,\BrandtBW\BilloFD\BilloDA-\LohmayerAJ}. They contain extra boundary terms in their effective action, and we will not discuss them here.} in dimension higher than three, $D>3$. The main result is that for closed strings there are no deviations from the formula~\energylevelsi\ before the order $1/L^5$. Most interestingly, the deviation from~\energylevelsi\ at the order $1/L^5$ is {\it  model-independent}, and proportional to $(D-26)$.
Excited states cease to obey~\energylevelsi\ at the order $1/L^5$, but the deviation at order $1/L^5$ of the ground state energy happens to vanish. (We do expect deviations starting at order $1/L^7$.)
The precise form of the energy levels of excited states at order $1/L^5$ was given in \AharonyGA; for example, the deviation from \energylevelsi\ for the first closed string excited states (with one left-moving and one right-moving excitation on the worldsheet) is given by $\Delta E_{1,1} = 4 \pi^3 (D-26) \alpha / 3 T^2 L^5$, where $\alpha$ depends on the $SO(D-2)$-representation of the state; $\alpha=D-3$ for the singlet, $\alpha=-1$ for the symmetric tensor representation, and $\alpha=1$ for the anti-symmetric tensor representation. While the deviations at order $1/L^5$ are universal, beyond this order the coefficients $a^{(k)}_{n}$ are expected to be model dependent.

It would be fascinating to verify these claims on the lattice. This requires precision measurements of the energies of excited states of strings for $D \geq 4$, which are technically challenging but should hopefully be possible in the near future.

For $D=3$
there are no deviations from~\energylevelsi\ at the order $1/L^5$. This prediction should be verifiable on the lattice.
We do expect deviations at higher orders, but we do not see any reason for them to be universal.

We focus on the computation of the energy levels, but there are also other universal properties of long strings, such as the fluctuation-induced width of open strings, which (for large $L$) can also be computed using the same effective string actions that we use here (for recent results see \refs{\GliozziZV\GliozziZT-\GliozziJH}). The effective action can also be used to study energy levels of high-spin mesons (with or without heavy quarks at the end), but we will not discuss this here.

The main tool we use is effective field theory. Since there is a gap for particles to go into the bulk, the low energy description should be that of the Nambu-Goldstone bosons of $ISO(D-1,1)/(SO(D-2)\times ISO(1,1))$, living in a two-dimensional space-time. There are various ways to present this theory, and we will discuss three different presentations in detail in this paper. In all these presentations we find that the leading order terms in the effective action (and, thus, also in the energy levels) are universal, while higher order terms are not. A priori it is not obvious that all these formalisms are identical, but at the end of the paper we present an argument showing that they can all be formally derived from the same underlying (string) theory, and thus they are equivalent (at least when some subset of non-universal couplings is involved).

The first formalism stems from the fact that there are no preferred coordinates on the worldsheet of a string-like object in a field theory. The idea is to keep this manifest by writing
effective actions that are coordinate-reparameterization invariant. This formalism is not very useful for computations, but it is useful for understanding the general structure of the effective action. This formalism enjoys some gauge redundancy, and it is not manifestly unitary.

 A second formalism \refs{\LuscherFR\LuscherAC-\LW,\AK} assumes
 that the string spans the coordinates $x^0,x^1$ (for long strings we can limit ourselves to such configurations, at least locally). The massless fields on the worldsheet can then be chosen to be $X^i(x^0,x^1)$ for $i=2,\cdots,D-1$.  The fields $X^i(x^0,x^1)$ describe physical displacements of the string into the directions $x^i$, and they are the only degrees of freedom in this formalism; they are NGBs for the broken translation symmetries.
 The effective field theory
\eqn\ftstatic{\int d^2x\CL[X^i(x)]~,}
 inherits from the space-time theory the shift symmetry $X^i\rightarrow X^i+const$. Therefore, the Lagrangian density only depends on derivatives of $X^i$. In addition, there is a nonlinear symmetry following from the fact that we can rotate or boost the string in the $i$'th direction. We call this {\it manifestly unitary} formalism the ``static gauge'' or ``unitary gauge'' formalism interchangeably; it can be thought of as a specific gauge-fixing of the reparameterization-invariant formalism. This formalism is useful for computations, and it makes many of the properties of the theory manifest, but other properties (like Lorentz invariance in space-time) are obscured.

A third formalism, first suggested by Polchinski and Strominger in~\PS, is to not commit on where the string sits, and to write an effective action for all the coordinates $X^\mu(\sigma^0,\sigma^1)$. However, one does not insist on reparameterization invariance of $\sigma^0,\sigma^1$. Instead, one insists on a symmetry under conformal transformations of the world-sheet, and in addition imposes algebraic constraints (similar to the Virasoro constraints in string theory).
This theory is not manifestly unitary. We refer to this formalism as the PS formalism. It is not clear how to directly derive this formalism from a gauge-fixing of the reparameterization-invariant formalism, but it passes many consistency checks and agrees with our results in all cases we checked; at the end of our paper we present an argument
 for the equivalence of this formalism to the other formalisms we discuss (at least up to the order we work in).
Because one has to work with a constrained set of variables, the quantization in this formalism is usually more complicated, and various facts that are easily visible in the other approaches are implicit in the PS formalism and, indeed, were not appreciated for a long time. The fact that the unitary formalism makes some of the salient (and measurable) properties of the string manifest, is another motivation for us to revisit this problem of effective string theory.

Various other formalisms have also been used in the literature to study effective strings~(see \refs{\PohlmeyerSQ\BotelhoZW\MeusburgerSA\ThiemannQU\MakeenkoDQ\MakeenkoUG-\BahnsWJ} and references therein), but we will not discuss them here.

In this paper we discuss energy levels purely within the $1/L$ expansion. As can be seen, for instance, from the Taylor expansion of \energylevelsi, this expansion is good only for $L \gg 1/\sqrt{T}$, and has a finite radius of convergence. Lattice results show that in many cases the deviations from \energylevelsi\ are small even where the power series in $1/L$ does not converge. An interesting possible explanation for this was recently given in \newNYU. They suggested a new method to compute the energy levels, by first computing the worldsheet S-matrix perturbatively in the derivative expansion using the effective action approach, and then using a Bethe ansatz-type approach to compute the energy levels, and they argued that this allows going to smaller values of $L$. It would be interesting to study this method systematically, and to understand if universal results can indeed be obtained even for relatively small values of $L$.

The outline of this paper is as follows. In section 2 we present the coordinate-reparameterization-invariant formalism, arguing for universality of the low-order terms in the effective action. In section 3 we discuss the static gauge (unitary) formalism. In both formalisms we conclude that up to (and including) the order $1/L^5$ there can be no model-dependent parameters in the energy levels of long closed strings. The correct answer deviates (for $D>3$) from the unjustified ansatz~\energylevelsi\ starting at order $1/L^5$, and at this order the deviation is model independent.
 This explains why the existing lattice results, within the range of convergence of the expansion in powers of $1/L$, are well represented by~\energylevelsi. In addition, it predicts that with better precision specific deviations will be uncovered for excited states in $D=4$.
 To illustrate our general results, in section~4 we discuss examples coming from confining strings in field theories with a gravitational dual. In section~5 we review the PS formalism, and argue for its compatibility with the other formalisms.

\newsec{Lagrangians for the Confining String}

Consider the worldsheet of a closed string-like object (for instance a flux tube) in $D$ dimensions; this is some
two-dimensional manifold $\CM$. The string lives in a
$D$-dimensional space-time with a flat metric. It can thus be described
by a mapping $X^\mu: \CM\rightarrow {\IR}^D$ ($\mu=0,\cdots,D-1$). We assume that any other degrees of freedom on the string worldsheet, including any excitations of the transverse shape of the string, are massive and can be integrated out in the low-energy effective action. The symmetry principles for the effective action $S[X^\mu]$ are
Poincar\'e invariance
acting on $X^\mu$, and independence of the coordinates chosen on
$\CM$. In the low-energy effective action we may assume that $\CM$ is smooth. Note that the space-time $\IR^D$ has a natural metric associated to it (the flat metric), while the string world-sheet $\CM$ is {\it not equipped with an inherent metric}.
Before attempting to write actions satisfying all these properties, we recall some elementary concepts from embedded geometry.

\subsec{Embedded Surfaces}

In this subsection, to simplify the discussion, we imagine that all the manifolds are Euclidean. Translation invariance in space-time immediately implies that we should
only consider terms in the effective action where $X^\mu$ appears with derivatives. There are several natural geometric objects to consider.
First of all, a metric on $\CM$ is induced from the flat metric in $\IR^D$
\eqn\inducedmetric{h_{ab}=\del_a X^\mu\del_b X_\mu~,\qquad (a,b=0,1).} This is
$ISO(D-1,1)$-invariant, and transforms under reparameterizations of $\CM$
like a two-dimensional metric. $h_{ab}$ is sometimes referred to as the first fundamental form.

In addition to the induced metric, we need to consider another object. At every point on the worldsheet we have two tangent vectors $\del_a X^{\mu}$ ($a=0,1$), and $D-2$ vectors orthogonal to them.
We can choose an orthonormal basis for these vectors,  $n_A^\mu$ with $A=2,\cdots, D-1$. Define a covariant derivative with respect to $h_{ab}$, $\nabla$, and differentiate both sides of~\inducedmetric. We find that
$\nabla_c \del_a X^\mu$ is orthogonal to the tangent space of $\CM$. Hence, there are functions $\Omega_{ac;A}$ such that
\eqn\defsec{\Omega_{ac}^{\mu}\equiv\nabla_c \del_a X^\mu=\sum_{A}\Omega_{ac ;  A} n_A^\mu~.}
$\Omega$ is called the second fundamental form. It is manifestly symmetric in $a,c$.
A certain quadratic in the second fundamental form reduces to the Riemann tensor via the equation of Gauss and Codazzi
\eqn\GCA{\Omega_{ac;A}\Omega_{bd;A}-\Omega_{ad;A}\Omega_{bc;A}= R_{abcd}~.}

Additionally, we can study derivatives of the normals $n_A^\mu$. We define
\eqn\defmu{n_A\cdot \del_a n_B=\mu_{AB; a}~.}
Orthonormality, $n_A\cdot n_B= \delta_{AB}$, implies that $\mu$ is anti-symmetric  $\mu_{AB; a}=-\mu_{BA;a}$. Note that $\mu_{AB;a}$ is non-trivial only for $D>3$. We can use \defmu\ to expand  $\del_a n_A^\mu$ in the basis $\{\del_aX^\mu,n_A^\mu\}$. We find
\eqn\normproji{\del_an_A^\mu=-h^{bc}\Omega_{ac;A}\del_bX^\mu-\mu_{AB;a}n_B^\mu~.}
As we remarked above, in three dimensions the last term is absent and thus the derivative of the normal can be expressed in terms of the first and second fundamental forms.

Let us discuss the transformation laws under change of basis of the normal bundle. We can perform an arbitrary orthogonal $SO(D-2)$ transformation (that depends on the point on $\CM$)  of the normals. The normals themselves, and the second fundamental form, transform homogeneously. However, the transformation law of $\mu_{AB;a}$ is
\eqn\mutran{\mu_a\rightarrow S\mu_a S^T+S\del_aS^T~,}
with $S \in SO(D-2)$. Thus, $\mu$ transforms as a connection of the normal bundle.

For our purposes, objects which are invariant under the structure group of the normal bundle, $SO(D-2)$,  can be expressed using derivatives of the tangent vector fields. Hence, they are expressible using the first and second fundamental forms and their derivatives.

We can define covariant derivatives using the connection~\mutran. We denote this covariant derivative by $\tilde\nabla$. For the normal vectors we get (using~\normproji)
\eqn\normal{(\tilde\nabla_a n)_A^{\mu} \equiv \left(\del_a\delta_{AB}+\mu_{AB;a}\right)n_B^\mu=-h^{bc}\Omega_{ac;A}\del_bX^\mu~. }
Hence, the covariant derivatives of the normals can be expressed using the tangent vectors, and the first and second fundamental forms. Another example is the field strength of the connection~\mutran, which one can easily compute to be
\eqn\fieldstrength{(F_{ab})_{AB} \equiv (\tilde\nabla_a\mu_b-\tilde\nabla_b\mu_a)_{AB} =h^{cd}\left(\Omega_{ac;A}\Omega_{bd;B}-\Omega_{bc;A}\Omega_{ad;B}\right)~.}
 This field strength transforms homogeneously under coordinate transformations and under reparameterizations of the tangent bundle. We see that it is also expressible in terms of the first and second fundamental forms, which both depend only on the tangent vector fields.

Thus, when writing actions, one loses nothing by considering only the tangent vectors and their derivatives.

\subsec{Reparameterization-Invariant Lagrangians}

We now apply what we have learned in the previous subsection, and write invariant Lagrangians. Effective actions are organized in terms of a
derivative expansion. Here the natural assignment is that the induced metric  $h$ carries weight zero, and any additional derivative carries weight one. Hence, the second fundamental form carries weight one.\foot{What we call ``weight'' in this paper was called ``twist'' in some previous papers on this topic.}

At weight 0 there is a unique admissible action, known as  the Nambu-Goto (NG) action
\eqn\mostgen{S_{NG}=-T\int d^2\sigma \sqrt{-h}~,}
where $h \equiv \det(h_{ab})$. This measures the area swept out by the world-sheet in space-time.
We assume that our string has a non-zero tension, $T$, so this will always be the leading term in our low-energy expansion.
The equation of motion of \mostgen\ can be nicely written with the first and second fundamental forms as
\eqn\eom{h^{ab}\Omega_{ab;A}=0~.}
Indeed, the equation of motion of $X^{\mu}$ gives $\square X^\mu=0$, which, according to~\defsec,  is equivalent to~\eom.

We need to understand the higher-weight corrections to~\mostgen.
Obviously there are no  corrections of weight one, so we immediately jump to weight two.
The only possible local terms of weight two are\foot{We only discuss local terms here. Since our formalism includes also non-physical fields it is not obvious that this is sufficient, but the inclusion of only local terms yields a consistent physical picture. }
\eqn\weighttwo{S_{(2)}=\alpha\int\sqrt{-h} (h^{ab}\Omega_{ab;A})^2+\beta \int\sqrt{-h} h^{ac}h^{bd}\Omega_{ab;A}\Omega_{cd;A}~.}
Using~\GCA\ we see that a combination of these two terms gives the Ricci scalar. Hence, we can rewrite~\weighttwo\ as
\eqn\weighttwon{\alpha\int\sqrt{-h} (h^{ab}\Omega_{ab;A})^2+\beta'\int\sqrt{-h} R~,}
where $R$ is the Ricci scalar of the induced metric $h$. The term proportional to the Ricci scalar is topological in two dimensions and since we will be expanding around long heavy strings that do not fluctuate too much, the Ricci scalar term plays no role. The term proportional to $\alpha$ is zero by the equation of motion of the weight zero Lagrangian~\eom, hence, it is a redundant operator that should not be considered.\foot{More precisely, terms which are proportional to the {\it exact} equations of motion are redundant operators that can be ignored. In the context of our expansion by weight, this means that terms proportional to the equations of motion of the lowest weight action~\eom\ can always be replaced by terms with higher weight.}
Incidentally, a specific combination of the terms in \weighttwo\ is known as the ``rigidity term'' \refs{\PolyakovCS,\KavalovDI}, which is also sometimes written (up to total derivatives) as $\int \sqrt{-h} (\square X^{\mu})^2$.
This term is interesting in the context of worldsheet actions which would be valid at all energy scales, since it is asymptotically free \PolyakovCS. However, in the context of our low-energy effective action this term is trivial at weight two, for the reasons explained above.

We conclude that there are no non-trivial terms at weight two. This is a surprising and important conclusion. To appreciate its significance, let us contrast this situation with several other theories of Nambu-Goldstone bosons.

In pion physics there is the universal two-derivative Wess-Zumino Lagrangian (roughly speaking, analogous to the NG action~\mostgen). This is followed at four derivatives by some nontrivial corrections that were classified in~\refs{\GasserYG,\GasserGG}. The four-derivative terms introduce new coefficients that need to be measured before predictions can be made about the four-derivative order. Hence, pion physics is {\it not completely predictive} at the next to leading order before a dozen parameters are measured.

Let us now compare the situation we are finding here with the spontaneous breaking of conformal symmetry~\refs{\SchwimmerZA,\KomargodskiVJ}. There, again, one has a universal term analogous to~\mostgen, but there are no terms obeying the Weyl symmetry at the next to leading order. Hence, the situation seems much more close to what we are finding here. However, in that case there is an anomaly term of the Wess-Zumino-Witten type at the next to leading order (and once we turn off the background metric it is a perfectly good term to write in flat space, invariant under conformal symmetry). This again introduces a coefficient that depends on the model. In the case at hand  we have found no such anomaly term. Hence, there is no new coefficient at the next to leading order.

At weight four there are several possible terms one could write, for instance, $\int \sqrt{-h} R^2$. So at weight four and higher, non-trivial terms exist, and we expect the effective action to be non-universal.

As we will explain below, this universality (the absence of counter-terms) at the next to leading order has important observable consequences.

\newsec{The Unitary Effective Action (Static Gauge)}

We now repeat the discussion of the previous section in a manifestly unitary setting, where there is no gauge redundancy. This is a convenient framework for computations. In addition, the analysis leads to several useful new points of view on this peculiar theory.

A natural choice of coordinates on $\CM$, for a string stretched mostly in the $X^1$ direction, is such that they coincide with $X^0$, $X^1$ in space-time. In other words, denoting the coordinates on $\CM$ by $\sigma=\sigma^1$, $\tau=\sigma^0$, we fix $\tau = X^0$ and $\sigma = X^1$. This gauge choice is possible for (and only for)
configurations which are sufficiently close
to a long static string stretched along the directions $X^{0,1}$. (For example, we do not allow the string to back-track or to intersect itself.)
The physical modes are then the transverse
fluctuations in the directions $X^i$ ($i=2,\cdots,D-1$), which are now functions of $\sigma$ and $\tau$. This choice fixes completely all the continuous gauge parameters (and does not lead to a nontrivial Faddeev-Popov determinant in the quantum theory).

One way to obtain allowed Lagrangians for the physical fields,
\eqn\genaction{\CL=\CL[X^i(\sigma,\tau)]~,}
 is by substituting the gauge choice above in the Lagrangians of the previous section. Here we will not rely on the previous section, and we will re-discover the properties of such long strings by directly considering the action of the physical modes.

When we write the action using the physical variables, only the $ISO(1,1)\times SO(D-2)$
subgroup of the $ISO(1,D-1)$ Poincar\'e group is linearly realized; the other generators are
non-linearly realized (since we have chosen particular directions for the string to lie in). The effective actions that we can write should be constrained not only by
the obvious $ISO(1,1)\times SO(D-2)$ symmetry, but also by the non-linearly realized
symmetries. The broken translations imply that the action can only depend on derivatives
of the $X^i$. The broken Lorentz transformations imply additional strong constraints.

Our effective action may naturally be expanded in powers of derivatives. If the
$X^i$ have dimensions of length (as is natural from their space-time interpretation),
then $\del_a X^i$ ($a=0,1$) is dimensionless. As in the discussion of the previous
section, it is natural to classify the terms in the action not just by the number
of derivatives, but also by their weight (which is defined to be the number of derivatives minus the number of $X^i$ fields). The notion of weight is useful because, as we will see, the broken Lorentz generators do not preserve the number of derivatives, but they do preserve the weight.
Translation invariance guarantees that all terms have non-negative weight, and the
$ISO(1,1)\times SO(D-~2)$ symmetry guarantees that all terms have even ``weight.''\foot{We
assume that our theories are parity invariant, both on the worldsheet and in space-time,
so that all terms must have an even number of $X$'s, and no $\epsilon$ symbols
appear in contractions of indices. There are also interesting non-parity-invariant
terms that can appear in the effective action~\refs{\BalachandranEP,\KavalovNX,\MazurNR}, but we will not
discuss them here.}

We begin by classifying all {\it weight zero} operators that can appear in the low-energy effective action
that are consistent with the linearly realized $ISO(1,1)\times SO(D-2)$ symmetry.
It will be convenient to use light-cone coordinates $\sigma^{\pm} = \sigma^0 \pm \sigma^1$.
Lorentz invariance on the string worldsheet implies that the weight zero terms
are polynomials in
\eqn\defyz{y \equiv \del_+ X \cdot \del_- X,\qquad\qquad z \equiv
(\del_+ X \cdot \del_+ X) (\del_- X \cdot \del_- X).}
(For the special case of $D=3$, $z=y^2$ and all weight zero terms are therefore polynomials in $y$.)
At second order in derivatives
we just have the standard kinetic term
\eqn\ltwo{{\cal L}_{2,2} = -{T\over 2} \eta^{ab} \del_a X \cdot \del_b X = 2 T y}
(whose normalization
we choose in a convenient way, fixing the normalization of the $X^i$; note that we use
here a different normalization than in \refs{\LW,\AK}). The subscripts of $\CL$ denote the number of derivatives and the number of $X$ fields, respectively. The weight is thus the difference between the values of the subscripts.

At fourth order in derivatives there are
two possible terms, $y^2$ and $z$ \LW.
At sixth order there are again two possible terms ($y^3$ and $yz$) \AK,
at eighth order three terms ($y^4$, $y^2 z$ and $z^2$), etc.

Terms of higher weight are a little more complicated to classify.
Setting to zero terms proportional to the equations of motion $\del_+ \del_- X^i + O(\del^4 X^3) = 0$ (such terms can be eliminated
by field redefinitions) and total derivatives, there are no such terms with four
derivatives or with two $X$'s,
and there is a single weight 2 term with six derivatives,
that may be written without loss of generality as
\eqn\lsixfour{{\cal L}_{6,4} = -32 c_4 (\del_+ \del_+ X \cdot \del_- \del_- X) (\del_+ X \cdot \del_- X)}
(this term is trivial for $D=3$).
At higher orders there is a large number of possible terms. The lowest order terms of weight four involve eight derivatives, for instance $(\del_+^2 X \cdot \del_-^2 X)^2$.

We have not yet imposed consistency under the broken Lorentz generators. There are several different methods to implement this.
Historically, this constraint was imposed by a very indirect method.
This method, which was initially proposed in~\LW\ and extended in~\AK, is to compute the partition function of the
effective action on an annulus or on a torus, to interpret this as a statistical
mechanical sum over closed string states (wrapped around a circle of circumference $L$) propagating over an open or closed interval (interpreted as a Euclidean time direction),
and to compare the result to the expected expression from a sum over strings with
energies $E_n(L)$. Lorentz invariance is used (as noted in~\MeyerQX) to rotate the Euclidean time direction
to the transverse dimensions when deriving this expected expression, and this leads to
constraints on the effective action. It was shown in~\refs{\LW,\AK} that up to six-derivative
order these constraints uniquely fix the values of the coefficients
%$c_2, c_3, c_6, c_7$
of the weight zero terms, but do not determine the value of the coefficient of the
weight two term $c_4$.
Extending this method to higher orders in the derivative expansion
is possible but technically complicated.

One of the main motivations for this work is to emphasize that a much more straightforward way of imposing this symmetry exists.  Consider, for example (without loss of generality), a Lorentz transformation
which is a rotation in the $X^1-X^2$ plane, $\delta X^1 = \epsilon X^2$, $\delta X^2 =
- \epsilon X^1$. One obvious change is the direct change in $X^2$,
specifically $\delta (X^2) = - \epsilon \sigma$ (recall that we chose the worldsheet
coordinates to be $\sigma=X^1$ and $\tau=X^0$). Additional changes arise because $X^1$
changes, which means that we need to change our definition of the $\sigma$ coordinate
to equal the new $X^1$ instead of the old $X^1$. This is simply a diffeomorphism
$\sigma \to \sigma + \epsilon X^2$. Taking both effects into account,\foot{It is very common that the implementation of some symmetries requires an accompanying gauge transformation. Here we needed a diffeomorphism transformation to accompany a boost in order for a boost to be a symmetry.
} we find
that under the $M_{12}$ Lorentz generator our fields transform as
\eqn\trans{\delta(\del_a X^j) = -\epsilon \delta_{a1} \delta^{j2} - \epsilon \del_1 X^j
\del_a X^2.}
The generalization to a general $M_{bi}$ ($b=0,1$,
$i=2,\cdots,D-1$) transformation is straightforward. We see that the transformation rule~\trans\ is indeed nonlinear, as expected.

The effective action must be invariant under these transformations; since the transformation
\trans\ leaves the weight of a given term invariant, we can analyze the possible invariant terms separately
for each value of the weight. In our analysis, we are mostly interested just in the leading correction
to the Nambu-Goto action, and we can use this to make several simplifications. First, we can drop
any terms that are proportional to the Nambu-Goto equations of motion. Second,
we can allow the transformation not to vanish but to be proportional to the Nambu-Goto equations of motion. This is because the effect of any term in the transformation that is proportional to the equations of motion is equivalent to the effect of a field redefinition, and we can always modify the Lorentz transformation law (at leading order) so as to cancel this field redefinition (note there is no guarantee that the classical transformation law \trans\ is exact).

The analysis of invariant terms is straightforward, so we just give the results here.\foot{An elegant method to perform this analysis was recently suggested in \refs{\GliozziCX,\MeineriEW}.}
For the weight zero terms there is a {\it unique} invariant action \refs{\JaimungalHK,\Aharony,\GliozziHJ,\GliozziCX}, which is simply the Nambu-Goto action written in the
static gauge,
\eqn\ngstat{\eqalign{S&=- T \int d^2\sigma \sqrt{-\det(\eta_{ab} + \del_a X^i \del_b X^i)}
\cr&=-T\int d^2\sigma\sqrt{(1-\del_0
X^i\del_0 X^i)(1+\del_1 X^j\del_1 X^j)+(\del_0
X^i\del_1X^i)^2}\cr&=T\int d^2\sigma\left(-1-\half
\del_aX^i\del^aX^i+\cdots\right)~.}}
%(See also \refs{\JaimungalHK,\Aharony,\GliozziHJ,\GliozziCX}.)
The uniqueness of the answer at weight zero is of course in agreement with what we found in the coordinate-independent framework.

The constant term represents
the tension of the string. The $X^i$ particles are approximately
free fields. If we take the string to be of length $L$, then clearly the energy levels due to the interactions in~\ngstat\ scale like
$E=TL+a_1/L+a_2/(TL^3)+\cdots$, where the $1/L$ term is the well-known L\"uscher term \LuscherAC. Note that \ngstat\ fixes the values of all the weight zero coefficients in the long string expansion.
The results agree with~\refs{\LW,\AK} but do not require any indirect considerations involving partition functions etc.

The action \ngstat\ is not renormalizable. Therefore one cannot simply use it to compute the energy levels without understanding the possible higher derivative (higher weight) counter-terms. Equivalently, in order to compute the complete series in $1/L$~\energylevel\  one needs to first specify infinitely many counter-terms (though at any given order in $1/L$, only a finite number of counter-terms contribute). Note that the situation is reminiscent of the energy expansion in pion physics (where there are infinitely many counter-terms beyond the two-derivative action, but to every order in the energy expansion there are just finitely many counter-terms, thereby rendering the theory useful).
This is a favorable situation, because it means that the $1/L$ expansion is predictive. For example, there are infinitely many energy levels, and to any given order in $1/L$ the infinitely many energy levels are determined by finitely many coefficients.

Hence, to be able to say something useful about the theory, we must investigate the higher-weight corrections.
At weight two, for $D=3$ there are no non-trivial terms, and for $D \geq 4$ there is no term that can be written whose variation under \trans\ vanishes exactly
\refs{\AD,\NYU,\GliozziCX}.
However, when $D\geq 4$, it is possible to write down a
{\it unique} weight two term that is invariant under \trans\ up to the equations of motion of the weight zero theory, and this is enough to guarantee the existence of a conserved charge associated with
Lorentz transformations at linear order in
this deformation. The derivative expansion of this term starts from the $c_4$ term~\lsixfour. For this term to be invariant, one needs to add to \lsixfour\ infinitely many other terms of weight two with more derivatives (and thus $X$ fields), all of whose coefficients are fixed in terms of $c_4$ \AD.

 It is important that the Lorentz variation of this weight-two chain does not vanish identically, but it is proportional to the equations of motion of the weight-zero theory. Indeed, applying \trans\ on \lsixfour, one finds a term with three $X$ fields, that vanishes only upon using $\square\ X^i=0$.
This means that for
the action to be invariant we must accompany the Lorentz transformation \trans\ by a specific
field redefinition, modifying the naive Lorentz transformation.
To leading order in $c_4$ the action is then invariant under this {\it modified} transformation.

However, $c_4$ is actually a {\it forbidden} counter-term in the action, for several reasons.
First of all, already at first order in $c_4$ we find that the algebra of Lorentz transformations is modified \NYU, which we do not expect. Indeed, consider two consecutive Lorentz transformations. We get a nonzero contribution when the
second transformation acts on the extra term coming from the field redefinition. More
specifically, consider a Lorentz transformation $M_{+i}$. At leading order in the derivative
expansion, this changes the term $c_4 (\del_+^2 X \cdot \del_-^2 X) (\del_+ X \cdot \del_- X)$
in the action by $c_4 \del_- X^i (\del_+^2 X \cdot \del_-^2 X)$, which by integration by parts
is equal to
\eqn\modtrans{-c_4 (\del_+ \del_- X^i (\del_+ X \cdot \del_-^2 X) + \del_- X^i (\del_+ X \cdot \del_-^2 \del_+ X)).}
This is proportional (again to leading order in the derivative expansion) to the
equations of motion, so it can be canceled by an extra term in the Lorentz transformation, of the
form $\delta_{+i} X^k = {c_4\over T} (\delta^{ik} \del_+ X \cdot \del_-^2 X - \del_+ X^k \del_-^2 X^i - \del_+ \del_- X^k \del_- X^i)$. Suppose we now consider acting first with $M_{+i}$ and then with $M_{+j}$; in the
joint transformation on $X^k$ we get an extra term (from the standard Lorentz transformation \trans\ generated
by $M_{+j}$)
\eqn\jointtrans{\delta_{+j} \delta_{+i} X^k = {c_4\over T} (\delta^{ik} \del_-^2 X^j - \delta^{jk} \del_-^2 X^i).} This term is
not symmetric in $(i,j)$, so it gives a non-zero contribution to the Lorentz commutator $[M_{+i},M_{+j}]$ which is supposed to vanish for $i\neq j$. Thus, when we add the $c_4$ term to the action, the Lorentz algebra is modified (at linear order in $c_4$ and at second order in the derivative expansion).
This suggests that the $c_4$ term should not appear in the effective theory of a string embedded in a Lorentz-invariant theory, since such a theory should have the usual Lorentz algebra.

The argument above is not completely convincing, since the string describes a field configuration that does not die off at infinity, so in principle it could modify the symmetry algebra. However, similar arguments imply that if we go to higher orders in $c_4$
the Lorentz symmetry will be broken (rather than deformed). The charge appearing in the leading-order Lorentz commutator $[M_{+i}, M_{+j}]$ described in \jointtrans\ is (at least classically) a conserved charge in the Nambu-Goto theory; this theory is integrable and has an infinite number of conserved charges. However, the theory with the $c_4$ deformation is no longer integrable, and the charge appearing in $[M_{+i},M_{+j}]$ is not conserved after this deformation.
So, even though the $c_4$ term preserves a deformed Lorentz symmetry at linearized order, it
does not seem to be compatible with such a deformed Lorentz invariance at higher orders.

Thus, there are no admissible terms in the effective action at weight two. This is the static gauge realization of what we found in the previous section. Hence, the results are consistent.

The conclusion that there are no weight two terms that can be written in unitary gauge means that there are no ambiguities at this level. Hence, the quantization of~\ngstat, {\it as long as one is careful to preserve the full Lorentz symmetry}, must give results that are trustworthy until the first order in the $1/L$ expansion that can be contaminated by some admissible weight-four terms. It is easy to see by dimensional analysis that weight-four terms cannot contribute before $1/L^7$. (This is because they start at the eight-derivative order.)
To summarize: {\it The results up to (and including) the order $1/L^5$ are completely universal, and can be computed by quantizing~\ngstat\ (being careful to preserve Lorentz)}.

The absence of counter-terms is a consequence of Lorentz invariance. In the quantum theory, it is sometimes convenient to introduce a regulator that breaks Lorentz invariance, and then one has to fix the answer by hand, by adding Lorentz non-invariant counter-terms such as~\lsixfour\ with a precise coefficient that tunes the breaking to zero \NYU.\foot{Other Lorentz-breaking terms are also induced, generally depending on the cutoff scale, and must also be subtracted.} We illustrate this by specific computations in the appendix.

One method which has been suggested in \ArvisFP\ (following \GoddardQH) for computing the energy levels is
a naive light-cone quantization of~\ngstat. This gives the ubiquitous formula~\energylevelsi.
There is no rigorous derivation of this quantization method from the Nambu-Goto action, and it clearly breaks space-time Lorentz invariance, since there is an anomaly in the Lorentz algebra for general $D\neq 26$, as reviewed in many textbooks, e.g.~\GreenSP. Hence, the result~\energylevelsi\ should not be trusted.\foot{An interesting proposal for a theory in $D<26$ for which \energylevelsi\ is exact was considered in the recent reference \DubovskyWK, but it is not related to the effective strings that we consider here.}
For the special case of $D=26$ there is a consistent fundamental string theory, and the result~\energylevelsi\ can then be rigorously derived by quantizing the Polyakov action of this string. Thus, in this special case this equation is known to be valid (for a specific choice of all the allowed higher-weight counter-terms). In the special case of $D=3$ there is also no Lorentz anomaly, so in this case the result~\energylevelsi\ seems to follow from a formalism that is consistent with all the symmetries (even though it has no rigorous derivation), suggesting that it may also be correct (again, for a specific choice of all the allowed higher-weight counter-terms), even though here we do not know any interacting fundamental string theory with these energy levels. However, even if ~\energylevelsi\ is a consistent spectrum in 3d, there is no reason to believe that it is exact for specific 3d theories like the pure Yang-Mils theory or the 3d Ising model. This is because there is no reason for these theories to have
precisely the choice of higher-weight counterterms leading to~\energylevelsi. Indeed, a theory in which the spectrum \energylevelsi\ is exact would develop a tachyon precisely at $L^2 = \pi(D-2)/3T$, while the theories in question develop tachyons at a slightly different radius (which can be interpreted as an inverse temperature).

As we will review below, computing the energy levels leads to answers coinciding
with~\energylevelsi\ for the orders $\CO(L),\CO(1/L),\CO(1/L^3)$, but deviating at order $1/L^5$ for excited states when $D>3$.
What we have learned so far is that this deviation at order $1/L^5$ cannot be ambiguous or depend on model-dependent coefficients; it will arise from any Lorentz-invariant theory that admits a long string.

\subsec{Energy levels of the effective string}

 Using the results above we can compute the energy levels of long effective strings, for instance when they are wrapped on a non-trivial circle of circumference $L$. As mentioned above, the leading term comes from the tension, going as $T L$, and the next term is the universal L\"uscher term proportional to $1/L$. Both of these terms come from the free part of the action \ngstat. The interactions in \ngstat\ then give (even before adding extra terms to the action) contributions to energy levels of order $1/L^{2n+1}$ for all integers $n$. Additional allowed terms in the effective action give similar contributions, starting at order $1/L^7$.

A direct computation at order $1/L^3$, at which the four-derivative terms in \ngstat\ contribute linearly, showed that equation \energylevelsi\ is correct to this order \refs{\LW,\AK}.
 At order $1/L^5$ one gets contributions from two insertions of these terms, which have not yet been computed. As we review below, the energy levels have been computed in a different formalism~\AFK, and these results suggest that at this order there should be a deviation from~\energylevelsi\ which is non-vanishing (for excited states with $D>3$) and proportional to $(D-26)$. It would be interesting to verify this by a direct computation in the unitary formalism~\newAK.\foot{An indirect argument for this result was given around equation 50 of \NYU, assuming that there is a relativistic theory on the worldsheet that gives rise to the spectrum \energylevelsi. Evidence for this assumption was given in \DubovskyWK.}

At order $1/L^5$, if one uses regularizations which require a $c_4$ term,
then this term also contributes to the energy levels. This is the case in particular for the zeta function regularization that is often used to compute the energy levels; this regularization breaks Lorentz invariance, and it was shown in \NYU\ that a specific value of $c_4$ was required in order for the Lorentz algebra to be satisfied. The contributions proportional to $c_4$ were computed explicitly in this regularization in \AharonyDB. Consistency requires that this regularization dependence exactly cancels with a regularization dependence in the computation of the contribution of the Nambu-Goto action to the energy levels at order $1/L^5$, so that the full energy levels are regularization-independent. It would be interesting to try to compute the energy levels using the Lorentz-preserving dimensional regularization, perhaps by relating them (along the lines of \refs{\DubovskyWK,\newNYU}) to the S-matrix in dimensional regularization.

Starting at order $1/L^7$, non-universal terms start contributing to the energy levels, and the leading contribution of these terms to the ground state energy was computed in~\AharonyDB. The effective string formalism is, however, still useful, as it implies that all energy levels at order $1/L^7$ can be expressed in terms of one or two unknown coefficients.

\newsec{The Effective Action for Long Holographic Strings}

As discussed in detail in \AK\ (see also \refs{\VyasWG,\VyasDG}), one case in which we can explicitly compute the effective
action on a long confining string is in gauge theories which are dual (by generalizations of
the AdS/CFT correspondence \refs{\MaldacenaRE,\GubserBC,\WittenQJ}) to weakly coupled
and weakly curved string theory backgrounds. In such backgrounds the long string is simply a fundamental (super)string moving in a ten-dimensional background, whose worldsheet theory includes the $X^{\mu}$ as well as $(10-D)$ additional scalar fields $Y^i$, and also
fermionic fields $\psi$. Several examples of such backgrounds are known (such as \refs{\WittenZW,\KlebanovHB,\MaldacenaYY}),
and it turns out that at leading order in the curvature, in the static gauge, they
all have the same action \AK, just with different values for the masses of the $Y$'s and
$\psi$'s. (We will discuss the action in another gauge in the next section.)

Naively one would expect all the additional scalars $Y^i$ to be massive, as the geometry localizes the string to a specific transverse position in these coordinates.
However, in all known examples, some of the $Y^i$ are massive with some mass $m$,
but some are massless and
parameterize a $p$-sphere which has some flux on it, so that it does not shrink to zero
in the IR region where the string lies. The additional scalars eventually develop a mass gap (since the sigma
model on $S^p$ has a mass gap), but the scale of this gap is exponentially small in the
weakly curved limit, and it cannot be seen in the perturbative expansion in the
curvature (the $\alpha'$ expansion in powers of $1/T$). One can, however, compute perturbatively the effective
action for the $X^i$ together with the classically massless $Y$'s, and this action is good above the
scale of the non-perturbative mass gap and below the scale of the other massive fields.
The fermions do generically acquire a mass, unless the background preserves some
supersymmetry, in which case some of them are massless Goldstinos (but we will not discuss
this case here).

In \AK, the effective action for the classically massless fields was computed at one-loop order
by integrating out the massive scalars and fermions. It was found that the $(\del X)^4$
and $(\del X)^6$ terms in the effective action precisely agree with Nambu-Goto (after
taking into account the renormalization of the tension~\BigazziZE), as expected from the general arguments
reviewed above. The $\del^6 X^4$ term (the term~\lsixfour\ in the static gauge formalism) can also be extracted from the
on-shell correlation function of four $X$ fields that was computed in \AK. On dimensional grounds one
can show that at one-loop order in the perturbative expansion of the string worldsheet theory, this term carries
no power of the mass, so it could either go as $\log(m)$ or be independent of the mass.
In~\AK\ it was shown that all the terms that depend logarithmically on the mass cancel,
so that the coefficient of this term is independent of the mass, and thus depends only
on the number of massive scalars and fermions.

The precise coefficient of this term
depends on the regularization. For example, in~\AK\ it was computed with a momentum cutoff regularization.
In this regularization, which breaks the space-time Lorentz symmetry, the contribution to $c_4$ from a massive scalar is $c_4^S = 1 / 192 \pi$
and from a massive fermion $c_4^F = 1/ 96 \pi$. In all the string backgrounds
in question the number of fermions $n_F = 8$, and if we denote the number of classically massless scalars by
$\tilde D$, then $n_S = 10 - {\tilde D}$, so we obtain $c_4 = (26 - {\tilde D}) / 192 \pi$ \Aharony.\foot{Note that the field theory corresponding to these backgrounds lives in $D$ dimensions, but we see here that there is a regime where there are in fact $\tilde D>D$ massless fields on the string worldsheet, with an approximate $ISO({\tilde D}-1,1)$ symmetry. This seems special to the holographic setup. Presumably, upon flowing to very low energies, where the sphere coordinates disappear and can be integrated out, the $c_4$ term will get corrected to $c_4 = (26 - D) / 192 \pi$. }

This result precisely agrees with the expected value in this regularization from pure field theory considerations, described in detail in the appendix.
Moreover, it is not difficult to repeat the same computation in dimensional regularization,
and in this regularization one finds a
vanishing contribution to $c_4$ from this computation, as expected. So, this example gives us a consistency check on
the general arguments of the previous sections and the appendix. In this framework it is not obvious that higher
loop corrections going as powers of $m^2/T$ do not modify the value of the $c_4$ term in the low-energy effective action, but the arguments above guarantee that this cannot happen, since the $c_4$ term must precisely cancel any Lorentz violations of the massless fields, and these are independent of $m$ (this also explains why the ${\rm log}(m)$ contributions to $c_4$ had to cancel in \AK).

\newsec{The Orthogonal Gauge and the Polchinski-Strominger Framework}

A third formalism for the effective action on long strings was suggested by Polchinski and Strominger \PS. Their formalism is motivated by fixing the diffeomorphism
symmetry on the worldsheet to the ``orthogonal gauge,'' in which the induced metric on the
worldsheet is proportional to the identity matrix, $h_{ab} = e^{\phi} \eta_{ab}$.\foot{One can always go to this gauge,
at least locally, by diffeomorphisms; we will not discuss global issues here.}
 One way to write this gauge-fixing condition, using
light-cone coordinates on the worldsheet, is $h_{++} = h_{--} = 0$. In this gauge it is convenient
to use dimensionless worldsheet coordinates, in which the length of the $\sigma$ coordinate is $2\pi$.

This gauge has several advantages compared to the static gauge discussed above. It manifestly
preserves the Lorentz symmetry in space-time. And, it preserves a larger subgroup of the
diffeomorphism symmetry than the static gauge, which is a conformal symmetry of the form
$\sigma^- \to f^-(\sigma^-)$, $\sigma^+ \to f^+(\sigma^+)$ (the resulting algebra
contains two copies of the Virasoro algebra). Another advantage of this
gauge is that the Nambu-Goto action becomes free in this gauge, since
\eqn\ngorth{S_{NG} = -T \int d^2\sigma \sqrt{-h} = -2 T  \int d^2\sigma h_{+-} =
- 2 T \int d^2\sigma \del_+ X^{\mu} \del_- X_{\mu}.}
The main disadvantages
of this gauge are that the description in this gauge involves non-physical degrees
of freedom (the longitudinal $X$'s and ghosts), and there are complicated constraints
on physical states related to the leftover gauge freedom.

Quantizing the theory in this gauge has some similarities to the quantization of a
fundamental string theory in the ``conformal gauge''; in a fundamental string theory
there is an independent worldsheet metric $g_{ab}$, and an extra Weyl symmetry acting
on $g$ in addition to the diffeomorphism symmetries, and using all of these symmetries
one can fix $g_{ab} = \eta_{ab}$, such that the Polyakov action \PolyakovRD\ becomes equivalent to \ngorth.
In particular, just like in the conformal gauge, if one implements the standard Faddeev-Popov procedure for gauge-fixing the
effective string to the orthogonal gauge,
one obtains a set of ghost fields $(b,c)$ with scaling dimensions $(2,-1)$ (both left-movers
and right-movers) which are decoupled from the scalar fields, and obey the Virasoro algebra
with a central charge $c=-26$. However, the BRST operator that
results from the gauge-fixing to the orthogonal gauge is naively quite different from the standard BRST operator of fundamental string theory.
Since the BRST operator associated with the orthogonal gauge is quite complicated, it is not clear how to
quantize the effective string theory directly
in the orthogonal gauge.

Polchinski and Strominger conjectured in~\PS\ that perhaps one can simply assume the same quantization rules as in fundamental
string theory, namely to impose Virasoro constraints on the physical states, requiring them to be
annihilated by (the negative modes in the Laurent
expansion of) the energy-momentum tensor. These constraints are self-consistent
if and only if the Virasoro central charge is $c=26$.

At first sight this leads to a contradiction,
since the Nambu-Goto action \ngorth\ in the orthogonal gauge leads to the Virasoro algebra with a central
charge $c=D$. But it was suggested in \PS\ that a specific correction to the effective string
action could cure this problem. Since the orthogonal gauge effective action is invariant under
the conformal algebra the Lagrangian must have scaling dimension $2$ (in units where $X$ is dimensionless), so one cannot write down
any other polynomials in $\del X$ in the action except for \ngorth. However, when expanding around
a classical solution corresponding to the long string configuration,
\eqn\expandlong{X^{\mu} = L(\delta^{\mu 0} \tau + \delta^{\mu 1} \sigma) + Y^{\mu} =
e_+^{\mu} L \sigma^+ + e_-^{\mu} L \sigma^- + Y^{\mu}~,}
(with $e_+^2 = e_-^2 = 0$ and  $e_+ \cdot e_- = -1/2$), we can allow in the long string effective action also terms which have negative powers of $Z \equiv -2 \del_+ X^{\mu} \del_- X_{\mu} = L^2 + O(Y)$.

The suggestion of~\PS\ is to add to~\ngorth\ a term proportional to
\eqn\sbeta{4\pi S_{\beta} = \int d^2\sigma \del_+ \log(Z) \del_- \log(Z) = \int d^2\sigma
{\del_+ Z \del_- Z \over Z^2} = \int d^2 \sigma {\del_+ \del_- Z \over Z}~.}
Note that this is reminiscent of the term introduced by~\PolyakovRD\ to describe non-critical fundamental string theory.
This term is  conformally invariant under the standard conformal transformation
$\delta X^{\mu} = \epsilon^-(\sigma^-) \del_- X^{\mu} + \epsilon^+(\sigma^+) \del_+ X^{\mu}$, as can be checked directly. One can show that this is the unique allowed term of weight two; some terms of higher weight were classified in \DrummondYP.\foot{Note that in \PS\ the weight two term \sbeta\ was written in a slightly different
form (see also \DrummondYP),
$$4\pi S_{\beta}' = \int d^2\sigma {\del_+^2 X^{\mu} \del_- X_{\mu} \del_-^2 X^{\nu}
\del_+ X_{\nu} \over Z^2} = \int d^2\sigma {\del_+^2 X^{\mu} \del_-^2 X_{\mu} \over Z}~.$$
This differs from the form above by terms proportional to the leading order (Nambu-Goto)
equation of motion of $X^{\mu}$, $\del_+ \del_- X^{\mu} = 0$. This means that the actions
$S_{NG} + \beta S_{\beta}$ and $S_{NG} + \beta S_{\beta}'$ lead to equivalent physics (up
to order $\beta^2$, or equivalently, up to higher weight operators),
but the relation between them involves a redefinition of $X^{\mu}$. This field
redefinition means in particular that for the action $S_{NG} + \beta S_{\beta}'$ to
be conformally invariant (even at leading order in $\beta$),
$X^{\mu}$ must transform in a more complicated way under
conformal transformations (as found in~\PS). We will work here with the simpler action
$S_{\beta}$, which is invariant under the standard conformal
transformation.}

The general action in the orthogonal gauge is thus $S_{NG} + \beta S_{\beta} + ({\rm higher\ weight\ terms})$ for some coefficient $\beta$. In the Polchinski-Strominger framework, this
action must lead to a Virasoro algebra with central charge $c=26$.
The conformal anomaly can be read off from the two-point function of the energy-momentum tensor $T_{--}$.
The conformal anomaly of the effective string action depends
only on $\beta$, and not on any higher weight terms; moreover, this dependence can only show
up at linear order in $\beta$ (these statements follow simply by counting powers of $L$).
It is not difficult to compute the conformal anomaly at this leading order in $\beta$,
as done in~\PS. Taking the action $S_{NG} + \beta S_{\beta}$, $T_{--}$ takes the form
\eqn\fortmm{
T_{--} = T_{--}^{NG} - {\beta\over L} e_+ \cdot \del_-^3 Y + O({1\over L^2}),}
and this corrects the conformal anomaly coming from the scalars to
$c_X = D + 12 \beta$. Thus, the effective action is conformally invariant if and only
if it contains the weight two term \sbeta\ with the precise coefficient \PS\
$\beta = (D - 26) / 12$. For any other value of $\beta$ there is a conformal anomaly,
so the constraints of the Polchinski-Strominger framework cannot be consistently imposed.\foot{Note that
$\beta$ is proportional to $(D-26)$. As we discuss in the appendix, in some regularizations the weight two term $c_4$ appearing in the static gauge is also proportional to $(D-26)$. However, generally there is no relation between these two terms.}
The fact that all terms
up to weight two are uniquely determined agrees with what we found in other formalisms
in the previous sections.

Since we do not know how to derive the Polchinski-Strominger framework directly by
gauge-fixing a diffeomorphism-invariant action,
we need to ask if this framework is
equivalent to the frameworks discussed in the previous sections, and in particular to the
static gauge effective action. One way to check this is by computing the S-matrix for the
physical massless modes in this framework, and comparing it to the one in the static gauge
(bearing in mind that the computation in static gauge should be done carefully, introducing $c_4$ to restore Lorentz invariance when needed). This comparison was done in~\NYU, and they
found a precise agreement. Another
comparison one can try to make is by computing the spectrum in the Polchinski-Strominger approach,
and comparing it to the spectrum of the static gauge discussed above. Up to order $1/L^3$ the static gauge
spectrum is simply the naive light-cone gauge spectrum \energylevelsi, and it was found in \refs{\PS,\DrummondYP} that the PS approach gives the same answer.
The spectrum to order $1/L^5$ in the PS formalism was computed in~\AFK, and was found to
differ from the naive light-cone gauge spectrum \energylevelsi\ by a universal piece (for $D > 3$) proportional to $(D-26)$. It would be interesting to confirm this in the static gauge.

We will now prove
the equivalence of the Polchinski-Strominger framework with the
static gauge approach (at least for some UV completions of the universal terms)
by using the long holographic strings discussed in the previous section.\foot{Our argument here implies that these formalisms are always equivalent up to the order we discuss in this paper, and that they are equivalent to all orders at least in theories, like gauge theories, which have a dual string theory description.}
We argued there that when we start with a fundamental string in a background dual to a confining
gauge theory, and go to the static gauge, the low-energy effective action becomes precisely of
the form discussed in section~3.
But we can also take this fundamental string action (explicitly written in \AK)
and fix it (using diffeomorphism
invariance and Weyl invariance) to the conformal gauge $g_{ab} = \eta_{ab}$. In this gauge
its symmetries and constraints precisely agree with the ones suggested by Polchinski and
Strominger.
We can now integrate out the worldsheet scalars corresponding to the
motion of the string in extra dimensions and the worldsheet fermions
(they are massive when expanding around the long
string solution), and this leads to higher weight terms in the action such as~\sbeta. Conformal invariance
guarantees that the coefficient $\beta$ comes out correctly;
for the bosonic string this was explicitly verified in~\Natsuume,\foot{In this case some of the extra dimensional scalars are necessarily tachyonic,
but this does not affect the result, and this problem does not arise for superstrings.}
and this can easily be generalized also to the superstring.

For these specific long strings we thus obtain the Polchinski-Strominger effective action (with their quantization rules)
and the static gauge effective action from one and the same starting point, proving that the two are equivalent.
It would be interesting to understand more directly how the orthogonal gauge gauge-fixing leads to
the term~\sbeta\ in the action and to the constraints suggested in~\PS.

\vfill\eject

\centerline{\bf Acknowledgements}

We are very grateful to A. Schwimmer for collaboration at various stages of this work, and for numerous useful comments. We would like to thank N. Berkovits, B. Bringoltz, M. Caselle, M. Dodelson, S. Dubovsky, M. Field, R. Flauger, F. Gliozzi, V. Gorbenko, M. Green, E. Karzbrun,  N. Klinghoffer, D.~Kutasov, J. Maldacena, V. P. Nair, J. Polchinski, N. Seiberg, M. Teper, A. Tseytlin, V. Vyas, and E. Witten for useful discussions.
OA would like to thank the Strings 2009 conference in Rome, and the ECT* workshop on ``Confining flux tubes and strings'' at Trento, for inviting him to present preliminary versions of these results. OA is the Samuel Sebba Professorial Chair of Pure and Applied Physics. The work of OA was supported in part by an Israel Science Foundation center for excellence grant, by the Israel--U.S.~Binational Science Foundation, by the German-Israeli Foundation (GIF) for Scientific Research and Development, by a grant from Rosa and Emilio Segre research award, and by the Minerva foundation with funding from the Federal German Ministry for Education and Research. OA gratefully acknowledges support from an IBM Einstein Fellowship at the Institute for Advanced Study.
ZK was supported by NSF grant PHY-0969448, a research grant from Peter and Patricia Gruber Awards,  a grant from Rosa and Emilio Segre research award, a grant from the Robert Rees Fund for Applied Research, and by the Israel Science Foundation under grant number~884/11.  ZK  would also like to thank the United States-Israel Binational Science Foundation (BSF) for support under grant number~2010/629.
Any opinions, findings, and conclusions or recommendations expressed in this
material are those of the authors and do not necessarily reflect the views of the funding agencies.

\appendix{A}{Regulator dependence in correlation functions in the static gauge}

In this appendix we explicitly analyze how the choice of regulator affects the value of $c_4$ in the effective action in the unitary formalism (static gauge), as discussed in section 3.

Consider the amputated on-shell four-point function of two $X^i$'s (with momenta $p_{1,2}
$) and two $X^j$'s (with momenta $p_{3,4}$), with
$i \neq j$ (this is only possible for $D > 3$, as we assume from here on).\foot{This on-shell four-point function is essentially the same as the S-matrix, but we use the four-point function because the definition of the S-matrix for massless particles in two dimensions is subtle, see \DubovskyWK\ for a nice discussion.}
The tree-level
Nambu-Goto action \ngstat\ contains a vertex proportional to
$\del_a X^i \del_a X^i \del_b X^j \del_b X^j - 2 \del_a X^i \del_b X^i \del_a X^j \del_b X^j$, which contributes to this correlator a term
proportional to $(p_1 \cdot p_2) (p_3 \cdot p_4) - (p_1 \cdot p_3) (p_2 \cdot p_4) - (p_1 \cdot p_4) (p_2 \cdot p_3)$. This is proportional to $s^2 - t^2 - u^2 = 2 t u$ where $s$, $t$ and $u$ are the
standard Mandelstam invariants $s = (p_1 + p_2)^2 = (p_3 + p_4)^2$, $t = (p_1 - p_3)^2 = (p_2 - p_4)^2$, $u = (p_1 - p_4)^2 = (p_2 - p_3)^2$, obeying $s+t+u=0$. In two-to-two scattering in two dimensions, it is always true that $t=0$ or $u=0$, so this correlator actually vanishes on-shell.

At one-loop order, there are several diagrams one can draw, using two copies of the vertex above (there is also a diagram with a six-point vertex from \ngstat). In one such
diagram we have a loop, with a scalar $X^k$ ($k\neq i,j$) running in it (there are $(D-4)$ such scalars).
It gives a correlator proportional to (denoting $p = p_1 + p_2$)
\eqn\oneloopamp{\eqalign{{\cal M}_{1-loop} = \int {d^2 k \over (2\pi)^2} {D-4 \over {k^2 (p-k)^2}}
& \left[((p_1 \cdot p_2) (k \cdot (p-k)) - (p_1 \cdot k) (p_2 \cdot (p-k)) - (p_1 \cdot (p-k)) (p_2 \cdot k)) \cdot \right. \cr
& \left. ((p_3 \cdot p_4) (k \cdot (p-k)) - (p_3 \cdot k) (p_4 \cdot (p-k)) - (p_3 \cdot (p-k)) (p_4 \cdot k)) \right].}}
This integral diverges quadratically, and we can evaluate it in various regularizations. Using momentum
conservation it is easy to see that the integral is equal to
\eqn\oneloopamptwo{{\cal M}_{1-loop} = (D-4) \int {d^2 k \over {(2\pi)^2}} {{((p_1\cdot p_2) k^2 - 2 (p_1 \cdot k) (p_2 \cdot k)) ((p_3 \cdot p_4) k^2 - 2 (p_3 \cdot k) (p_4 \cdot k))} \over {k^2 (p-k)^2}},}
or (by introducing the usual Feynman parameter) to
\eqn\oneloopampthree{{\cal M}_{1-loop} = (D-4) \int_0^1 dx \int {d^2 l \over {(2\pi)^2}} {{((p_1\cdot p_2) l^2 - 2 (p_1 \cdot l) (p_2 \cdot l)) ((p_3 \cdot p_4) l^2 - 2 (p_3 \cdot l) (p_4 \cdot l))} \over {(l^2 + x (1 - x) p^2)^2}},}
where we assumed the expression is regularized to exchange the order of integrations.
If we now perform the angular integrals, assuming for a moment that the integral over $l$ is in $d$
space-time dimensions, we find
\eqn\oneloopampfour{\eqalign{{\cal M}_{1-loop} = (D-4) & \int_0^1 dx \int {d^d l \over {(2\pi)^d}}
{l^4 \over {(l^2 + x (1 - x) p^2)^2}} \cdot \cr & \left[ \left(1 - {4\over d} + {4 \over d(d+2)}\right) (p_1 \cdot p_2)^2
+ {4\over d(d+2)} (p_1 \cdot p_3)^2 + {4\over d(d+2)} (p_1 \cdot p_4)^2 \right].}}

We can now consider various regularizations. In any regularization that leaves the integral two
dimensional, such as a cutoff regularization or a Pauli-Villars regularization, the term in the
square parenthesis in \oneloopampfour\ is proportional to $(s^2-t^2-u^2)$, just like the tree-level
contribution, and thus it vanishes on-shell. So, in these regularizations there is no one-loop
contribution to this correlator. On the other hand, in dimensional regularization, the
square parenthesis are proportional to (up to linear order in $(d-2)$) $[(s^2-t^2-u^2) - (d-2)(s^2+3t^2+3u^2)/4]$, which on-shell is the same as
$(2-d)s^2$. When this multiplies the remaining integral over $l$, which (near $d=2$) is proportional to $1/(d-2)$, one obtains a
finite answer in this regularization, which (in the $d\to 2$ limit) gives an amputated correlator equal to $[-(D-4) s^3/(192\pi T^2)]$. Similarly, the additional diagrams (with $X^i$ and $X^j$ running in the loop) also
give zero in two-dimensional regularizations, and in dimensional regularization they give a non-zero
answer which modifies the coefficient $(D-4)$ to $(D-26)$.\foot{The other non-trivial on-shell correlators are related to this one by crossing. Note that the correlator with $i=j$, which is the only one that can be computed for $D=3$, is trivial at this order in the derivative expansion, since it is proportional to $s^3+t^3+u^3=3stu$ which vanishes on-shell.}

Of course, different regularizations must eventually give the same answer, so how is this possible?
The expansion in loops in the theory \ngstat\ is the same as the derivative expansion (both are expansions
in the inverse tension), and
there is another possible contribution to the correlator at the same order, which comes from the
$c_4$ term \lsixfour\ at tree-level; this term gives an on-shell correlator going as
$[-c_4 s^3 / T^2]$. Thus, consistency requires that the value of $c_4$ in the effective
action depends on the regularization which we use, and that the value of $c_4$ in dimensional regularization differs from its value in two-dimensional regularizations by $(D-26)/(192\pi)$.

As discussed in section 3, in regularizations which preserve Lorentz invariance, such as dimensional regularization, the $c_4$ term in the effective action must vanish. The discussion above then implies that in any Lorentz-invariant theory, in which the effective string is regularized by a two dimensional regularization of the continuum momentum integrals\foot{Note that this argument does not determine
the $c_4$ term for regularizations like the zeta function regularization of the effective string
compactified on a circle, that were used in \AharonyDB, since these regularize momentum sums rather than the
continuous momentum integrals used above. This is consistent with the claim of \NYU\ that in the zeta
function regularization $c_4$ takes the value $c_4 = -1 / 8 \pi$.}, the $c_4$ term must take the value $c_4 = (26-D) / (192 \pi)$ \Aharony. This is confirmed by the explicit computations described in section 4. Presumably, one can alternatively fix $c_4$ by requiring that the Lorentz currents (to the extent that these are well-defined in the static gauge) are exactly conserved.

\listrefs

\end